# On the Nature of Time

Mario Radovan
University of Rijeka, Croatia

**Abstract**

The paper puts forward a conceptual framework in which the phenomenon of time can be presented and discussed in a proper way. We argue that change is ontologically and epistemologically a more basic phenomenon than time. Time is an abstract entity created by the human mind on the basis of the experience of change. Physical reality is a process of ceaseless becoming and vanishing; time is not a part of that process. Time is the abstract bank in relation to which we measure the intensity and amount of the flow (change) of physical reality. We must differentiate physical reality from abstract entities (language) by means of which we speak about this reality. It is necessary to differentiate a formal description (formulas) from its interpretation: a correct formal description can be interpreted in a logically inconsistent and factually wrong way. We argue that the discourse about the relativity of time joins (mixes) physical reality and language, and gives an inconsistent interpretation of correct formulas. Regarding the future of time, it has been said that physicists are divided between two options: (1) to pin down a "master time", as a measure of change of physical reality, and (2) to proclaim time "out of existence". We argue that *both* options must be adopted, because time is (1) a measure of change, created by the human mind, (2) and time is an abstract entity that does not exist in physical reality.

## 1. The river and the bank

1. "Time is one of the last great mysteries", runs the first sentence of a large anthology of texts about time (Callender 2011a, 1). Different physical theories often give different images of time, which are not mutually compatible, so that they have not resolved the mystery. It has been said that the awareness of the flow of time is the core element of the human understanding of the world and of human existence. Time has been called an invisible river that carries forward all things until they sink in its waves. But time is not a river and it does not flow. Physical reality is a *process* and, figuratively speaking, it can be called the *river of existence*. This river does not need time or any other force beyond itself to "carry" it and to be what it is: a *lasting process*. Time is not part of that process (river): it is not an ingredient of physical reality and it does not flow. Time is an *artificial bank*, created by the human mind, in relation to which the river of physical reality flows. Time is an element of our language by means of which we *speak* about physical reality and its changing: it is one of the basic elements of the conceptual system by means of which we express our perception and understanding of the changing physical reality in which we dwell and of which we are part. In sum, I do not see anything particularly mysterious about time.

2. Many people believe that wisdom is in formulas: if a text contains plenty of sophisticated formulas, then it surely contains something very wise and important. But this is not so: formulas are needed for computation, but they are not essential for the correctness of a scientific view and narrative. What matters the most are ontology (basic concepts) and consistency of discourse. It is necessary to understand and to present (describe, define) in a clear way what we speak about. Our discourse must be logically consistent to be understandable (even if wrong), and to have a chance to be correct. Many people use complex formulas with an aim to impress readers and cover up the fact that they do not have anything relevant to say. In his *Principia*, Isaac Newton presented the basic laws of physics without using formulas. A clear ontological (conceptual) framework and logical structure are the most important for a discourse that wants to explain something. Formulas come later, and they can wait.

3. The physical world is a river that flows: time is the measure of the amount and intensity of this flow. Physical entities are not "carried along" by time: they are simply *changing* by their own nature; things change, they are not carried anywhere. People express the experience of change in terms of time as a linguistic means. Time is an abstract dimension on which the human mind projects its experience of the changing reality. Change is immanent to the physical world. The passage of time and our passage in time are metaphorical expressions of the fact that everything changes. Time does not pass, but *we* do: we are the sandglasses that are not turned over when their sand runs from the upper part to the lower. Each of us is an hourglass that runs only once. We do not measure the passing of time: we measure our own passing and vanishing.

We define and measure time by means of various cyclic processes, but a cyclic process is not time: it is a process. Time is an abstract measure of the amount of change, expressed in terms of some chosen cyclic processes, such as the rotation of the earth around the sun and around itself, or the oscillation of certain atoms or other particles. When we say that a person is forty, we say how long the *process* that this person *is* has been going on, and we do this in terms of the cyclic movement of the earth around the sun. We speak by means of concepts (abstract entities), but we always compare physical processes.

4. Physical reality is a process of becoming and vanishing; the human mind projects its perception and experience of this process onto an abstract dimension called time. Points in this dimension are determined (set) by means of a cyclic process which we consider the most stable one we currently know, or sufficiently stable for certain purposes. Time is a creation of the human mind: it is a metaphysical (linguistic, cognitive) category, not an ingredient of the physical world. Time belongs to language, not to physical reality. The same holds for other basic elements of the conceptual system by means of which we express our experience and understanding of existence. Other core elements of that system are space and causality; numbers must be added to the list of core metaphysical categories in terms of which we think and speak. We need space, time, numbers, and causality as the means for describing our experience and understanding of physical reality. But this does not mean that these entities exist in the physical sense, beyond the space of abstract (linguistic) entities.

5. Time is "a fundamental characteristic of human experience", says James Whitrow. Such claims are not correct: we experience change, not time. "But there is no evidence that we have a special sense of time, as we have of sight, hearing, touch, taste, or smell", continues Whitrow (1989, p. 5). Robin Le Poidevin also points out that people have organs of sight, hearing, smell, and for receiving other inputs, but do not have a "sense-organ for time" (2009, 100). However, this is normal, because time is not an ingredient of the physical world to be sensed, so that there was no reason for such an organ to develop. Time, sight, and touch are quite different things and they should not be mixed together. It is not strange that people do not have a special sense of time, because time cannot be sensed in the way that light and sound can be; time is neither a part nor an attribute of the physical world.

6. I have difficulties with the discourse about *temporality*. "Macroscopic processes appear to be temporally 'directed' in some sense", says Craig Callender (2011b, 1). I do not see what would be lost if the qualification "temporally" were left out from the above sentence. What is the difference between the claim that physical reality is a process that seems "temporally directed" and the claim that physical reality is a process that seems "directed"? If a process is directed, then it moves away from some state, and closer to (towards) another state. One can call such directed movement temporal, but this does not mean that besides the movement (that is, a directed process) there exists something called *time* which affects ("directs") this movement. "Our world is filled with processes that have an evident time direction", says Mauldin (p. 166). The physical world (universe) is not "filled" with such processes: it *is* such a *directed* process. However, I do not see why "time" or "temporal" direction is needed here nor what this means. If we leave out the word "time" from Mauldin's claim, what will be lost? Nothing, as much as I can see.

7. If something is a directed process, then the claim that this is a "temporally" directed process does not tell anything new. A process can be described as directed or cyclic, perhaps. But to say that it is *temporally* directed seems a redundant addition to me. Discourse about time contains many questions and claims that I consider vague and without a clear meaning. Let me mention a couple of such examples. "What is the origin of the thermodynamic asymmetry in time?" (Callender 2011b, 1) Why is "in time" needed here? Asymmetry is a basic feature of the process called physical reality. This process is such as it is, and it is not "in" anything: it simply *is*, and it is exactly such as it is. "Does the thermodynamic time asymmetry explain the other temporal asymmetries? Does it account ... for the fact that we know more about the past than the future?" (Callender 2011b, 1). I am impressed and confused with such profound questions. My profane answer to the above question could run like this. The past is something that I or others have seen, so that we can know some things about it. The future is something that I and others have not seen, so that we do not know it. My profane answers are usually not considered good enough to match the profundity of questions like the ones we put forward above.

8. In sum, we argue that time does not exist in the physical world: it does not flow, because it is an abstract entity created by the human mind. Time and space are dimensions of the abstract framework in which we can speak about the changing physical entities, of their becoming and vanishing. Until physics accepts these basic ontological facts, its discourse about time will be in formal and factual trouble. Towards the end of his book, Paul Davies raises the essential question, which his book does not answer. "Suppose you met an alien who claimed he had no idea what you meant by the flow of time", he says. Well, you were lucky and you met one: me. How would you *describe* the flow of time to such an alien: "what would you say to convince him of its reality?" - asks Davies (p. 255). Indeed, what? Nothing, I am afraid, because there is nothing convincing to be said in this regard.

**2. A matrix of discourse**

9. Many have raised the question whether time is "real" or "unreal", whether it "exists" or it is only an "abstraction". Similar questions have been asked for numbers, particles, forces, and other things. Those who love to deal with such questions do usually not tell us what it *means* to exist, to be real or unreal. Hence, to make the discourse more precise, let us put forward brief descriptions of several basic concepts. We speak about the matrix of discourse, because we structured the selected concepts in three "columns", each of which consists of three concepts ("rows"), which together form a matrix. The first column of this matrix shows a basic ontological division of reality: we speak about (1) physical, (2) mental, and (3) abstract entities. The second column points at the difference between (1) reality in itself, (2) a formal (mathematical) description of reality, and (3) the interpretation of a formal description, which explains what the formal description actually says (means) in ordinary language. The third column of our matrix of discourse consists of three basic disciplines of our discourse about human thinking and knowledge: these are (1) ontology, (2) epistemology, and (3) logic. We can say that the first of the three columns deals with the nature of reality, the second column addresses the issue of knowledge, and the third column presents three levels of discourse.

10. We consider appropriate to introduce a basic ontological framework which facilitates and requires a division of "everything that is" (what we perceive, feel, or think about) into the following three classes: (1) *physical entities*, (2) *mental entities*, and (3) *abstract entities*. These three classes have been called "worlds", so this ontological framework has been called the three-world ontology. Let us denote the class (space or world) of physical entities with C1, the class of mental entities with C2, and the class of abstract entities with C3. Stones and rivers belong to C1, pleasures and pains belong to C2, numbers and languages belong to C3.
In the context of the given ontological framework, we argue that time is not a physical entity, like stones and rivers, and that it does not belong to the physical world (C1). Time is not a specific mental state either, such as the pain I feel in my left knee, and my love for Esmeralda, so that it does not belong to C2. Time is a creation of the human mind and it belongs to the space of abstract entities (C3). Time is an abstract means by which we measure (quantify) the amount of physical change and the intensity of this change (Radovan 2011; 2014). All measures are products of the human mind; therefore, they are abstract entities. Stones (C1), my feelings (C2), and time (C3) all exist and can be called real, but they dwell in different spaces of existence and reality.

11. Let us see the second column of our matrix of basic concepts. In our discourse about physical reality and knowledge, we must differentiate (1) the structure of physical reality in itself, (2) formal (mathematical) description of a class of phenomena, and (3) the interpretation of a formal description, which explains what a formal description actually means in common language. The structure of reality in itself is unreachable: it cannot be known and it seems inconceivable. It is not possible to know reality in itself, because knowledge is a *relational* category: knowledge is a "joint product" of the observed *object* and of the capacities, means (devices, language), and aims of the *subject* that observes the object. Our knowledge is never absolute or "pure": we always *shape* what we see: we do this with our capacities, means, and aims.

12. By the formal description of phenomena, we mean a description by means of formulas. A formula expresses correlations between certain attributes of a class of phenomena: it describes the behaviour (change) of an attribute (of a phenomenon of the given class) in terms of the behaviour (change) of some other attributes of that phenomenon. A formula tells us how the changes of values of certain attributes of a phenomenon influence (cause the change) of the value of some other attribute of that phenomenon. Formulas receive and produce quantities (numbers): they function like machines which receive quantities as inputs and produce quantities as outputs. We take the output (value) that a formula produced and compare this value with the value of the corresponding attribute of a phenomenon, obtained by measuring. If we are happy with the outcome of such a comparison, we say that the formula is correct and that the theory of which the formula is part is correct.

The claim that a formula is correct means that the results it produces have been corroborated by observations and tests. The correctness of a formula can be tested and corroborated by observations and experiments, but the correctness of a universal formula cannot be definitely proved, because it is always possible that some future test or discovery will show that the formula is *not* correct for a specific subclass of phenomena. In such case (that is, if this happens), we must make some change in the formula, or narrow the scope of its application, so as to exclude the critical subclass of phenomena. In general, a universal formula cannot be proved definitely correct by testing and observation, but testing can always prove that a formula is not (quite) correct.

13. Formulas are mute: they do not say what they, their inputs, and outputs *mean*. Hence, a formula is normally accompanied with explanation which says what the formula actually means or says in common (or informal) language. Such an explanation is an interpretation of the formula and a partial interpretation of the theory of which the formula is part. Formulas can be empirically tested, but interpretations cannot be tested in such a direct way. A formula can be interpreted in different ways, and a correct formula can be interpreted in a wrong way. Scientists test formulas, and when the results of tests confirm a formula, they normally take this as the confirmation of their *interpretation* of the formula, too. But this is not a logically valid step and it can lead to a wrong conclusion. In the next paragraph, we put forward an example of the interpretation of a formula, which is normally presented without mentioning the difference between facts and their interpretation.

14. Some particles which have short decay half-time (lifetime), have a very small rest mass, so that they can be accelerated to a very high speed. By means of Lorentz transformations we can compute the *increase of lifetime* of these particles when they move with the speed $v$ relative to the earth (laboratory). The results computed by these formulas have been empirically confirmed many times, so these formulas are considered correct. However, here ends the factual discourse: the *explanation* of this fact is a matter of interpretation. My interpretation of the increase of lifetime of the particles runs like this. Particles are *processes* (they decay): the fact that the increase of the speed of their movement is accompanied with the increase of their decay half-time (lifetime) indicates that with the increase of speed, *processes* in the physical world evolve more slowly. On the other hand, relativists interpret the same empirical fact in the following way: with the increase of speed with which an entity moves, *its time* flows more slowly.

We argue that the discourse about the slowing down of time is structurally wrong because it mixes ontological categories: processes, which are physical entities (C1), and time, which is an abstract entity (C3). We also argue that the standard discourse about the slowing down of time is logically inconsistent, which means that it does not explain anything. Relativists deny both: they argue that space and time "are *physical* things, mutable and malleable, and, no less than matter, subject to physical laws" (Davies, 16). They also deny that there is any contradiction in their discourse about the slowing down of time. Anyway, the correctness of formulas does not prove the correctness of a given *interpretation* of these formulas.

15. Let us mention that like time and numbers, causality is also a product of the human mind. Strictly speaking, events (and entities) do not cause each other, because there are no separate events and entities in themselves. Existence in itself is a shapeless *One* and it has no intrinsic divisions and borders. The human mind, with its perceptive and cognitive capacities, imposes a multiplicity of shapes to the shapeless One, and by this it creates entities and events. The human mind that scatters the One into a multiplicity, created the idea and concept of *causality* as the means by which it tries to connect together the multiplicity of entities it produced. By means of causality, the mind tries to connect entities it created, and to explain the functioning of the image of reality it produced.

16. Let us see the third and last column of our matrix of basic concepts. We can speak about knowledge and reality at three different levels: (1) ontological, (2) epistemological, and (3) logical. Ontology aims to establish what are the basic elements (or ingredients) that reality is made of, and what are the features of these elements. It is not possible to know reality in itself, so that ontology tries to find a set of basic *concepts* (categories) by means of which (and of those derivable from them) we can *describe* the reality we experience, in the most suitable (effective) way. Such *descriptive ontology* is hypothetical rather than assertive: it analyses concepts and conceptual systems, points at their defects (inconsistency, incompleteness, redundancy) and seeks better solutions.

Epistemology deals with the issue of knowledge: it studies the nature, origin, and limits of human knowledge: it tries to establish what can be considered knowledge and on what basis. Epistemology aims to establish the principles and methods on the basis of which we can decide which claims (statements) ought to be considered *knowledge* and which ought not be considered so. The main issue of epistemology regards the question of justification (testing, verification) of claims. It is often difficult to tell whether a justification (of a claim) should be considered correct (valid, strong enough), and it seems that no empirical justification can be absolutely valid (such that it cannot be confuted). We do not need to deal with these issues in more detail here.

17. Logic deals with the issues of validity of inference and of consistency of discourse. What makes logic particularly important is that it sets the threshold below which a discourse (theory, interpretation) *surely* contains false claims (statements) and loses a clear meaning: this threshold is consistency. A discourse (theory) is logically consistent if and only if it is not logically inconsistent: a discourse is logically inconsistent if and only if it contains or logically implies a statement *p* and its negation *not(p)*. In other words, an inconsistent discourse claims and negates the same thing (proposition), so that it surely contains at least one statement that is not correct (true). An inconsistent discourse does not express a clear meaning: it is not possible to tell what an inconsistent theory actually says, because it negates what it claims. An inconsistent discourse (interpretation) cannot be a correct description of anything.

Discourse in the physics literature often shows a lack of understanding of logical concepts. For example, speaking about the issue of travelling through time, Paul Davies says that "the laws of the universe must by definition describe a consistent reality" (p. 249). A better way of explaining the relationship between laws (discourse) and physical reality could run like this. Physical reality *is* such as it is: it makes no sense to call it either consistent or inconsistent; consistency regards *conceptual* systems (theories, discourse), not physical (material) systems. A theory as a set of statements (definitions, laws, formulas) must be logically consistent to be understandable (meaningful, even if wrong) and to have a *chance* to describe an observed reality in a correct way. An inconsistent theory necessarily implies factually wrong claims, it does not express a clear content (meaning), and cannot describe any reality in a correct way.

## 3. The origin of time

18. For a discourse about time, we need the concepts of past and future. For the past and future to exist, we need the present ("now") which stands between them and in relation to which they *are* the past and the future. "Now" is a *state of mind* and it exists only for the conscious mind. There is no "now" (present) in the inanimate physical world, and physical theories have no means (concepts) by which they could express such a state. This means that there is "no future or past, and no arrow or flow of time in the inanimate world", says Julius Fraser (p. 242). It should be mentioned here that Fraser speaks about a *hierarchy* of different temporalities (times), but we do not need to deal with the details of his discourse here, because it does not change the essence of the problem. Physics speaks about the "now-less" physical reality (C1): "now" is a subjective state (C2) and it transcends the language of physics. The inanimate physical world is "now-less", and consequently, it must be considered time-less in itself. Time is not a cosmic phenomenon, ontologically independent of the mind. Events in the inanimate physical world can be considered "ordered" or "directed" in some way, but there is no basis for considering this order *temporal* in itself.

19. Like time, mind has been called one of the "last mysteries" that transcends the reach (language) of contemporary natural sciences. Mind has been a topic of intense philosophical and scientific debates, but these debates have not produced a satisfactory (useful) theory of the mind.

Hence, I use here the concept "fissure" as a figurative description of the relationship between the conscious mind and the physical reality from which it emerges. Each consciousness is a *fissure* on the face of physical reality, in which (fissure) the reality (as a process) reflects itself. Consciousness is not a new kind of substance (C1): it is a structural property of substance, which bestowed *quality* upon existence. Consciousness opened the space C2 and facilitated the creation of the spaces C3.

"Now" (the present) does not move (advance): physical reality changes ("flows") in front of the conscious mind ("fissure") that observes this reality. The world is a process; the present is the current image of this process, captured by the "camera" of the human mind, called consciousness. This is a figurative discourse, but I know no better description of the relationship between physical reality and conscious mind. What matters for us here is that the present ("now") is not a feature of the physical reality in itself. For a "now", consciousness is necessary.

Time is born with the appearance of consciousness. Without a conscious mind, events on the earth, the activities of living beings included, would evolve as *processes*, as they were evolving before the conscious mind appeared on its surface. We can describe such a world as a *reality in motion and change*. However, there is no *time* in such a world, because it is not possible to speak about *time* as something independent of the conscious mind. This is a conceptual issue, not empirical: clocks do not know time; only a conscious mind can know time. The mind can project its knowledge of time on the inanimate physical world, but this is a matter of the interpretation of reality, rather than of reality itself.

20. There are claims that the passage of time is necessary for changes to take place. The relationship between time and change seems problematic, but this is essentially a consequence of the wrong approach to this issue. It is wrong to start from the position that events take place "in time" and that change "needs time" to take place. Change is *immanent* to physical reality; change is also the basic element of the human perception and understanding of that reality. Physical reality is a process of ceaseless becoming and vanishing. Change is *ontologically and epistemologically prior* to time: we perceive change, not time. If there were no change, nobody would speak about time: people would have not *invented* it. In sum, time is an abstract means by which the mind describes its perception and understanding of reality as a process and change. Time is a means of discourse: it does not exist beyond the human mind and language, except in the realm of abstract entities (C3) created by the human mind and language.

Change is immanent to physical reality: it simply *is*, without any preconditions. There is nothing "inferential" in the perception of change: "it is simply given in the experience", says Robin Le Poidevin (2009, 87). On the other hand, time is *created* by the mind from the experience of the *directed change* (C1). Time is derived from change: it is an abstract dimension (C3) onto which the mind expresses (projects) its experience of the changing reality (C1). There are claims that "we perceive time through motion" (Davies, 29), but such claims are wrong. We do not perceive time through anything, because there is no time in the physical world to be perceived. We perceive *change:* on the basis of this experience, we created time as one dimension of the space of discourse about reality, onto which we express our experience of change.

21. To say that change takes place "in time" is superfluous and it means nothing. Change is immanent to physical reality and to our perception of this reality. Change does not need anything more basic than itself to explain itself: it simply *is*. *I am who I am*, said Yahweh god; change can say nearly the same: I am what I am, and I am the source of everything there is. Every explanation must stop at some point. We do not need anything more basic than change to describe physical reality, and there is nothing more basic than change, which could explain it. Change is ontologically and epistemologically prior to time, and change does not need it. Time has been introduced for practical reasons and purposes, but it does not give a deeper explanation of reality. Change is the source of everything else, and it must be the basis of every explanation and discourse.

22. We cannot present here the views and reflections about time of various outstanding thinkers throughout history. Let us only mention a couple of basic things. Aristotle's discourse about time is very interesting and it seems correct. He basically reduces time to "a measure of motion", but his discourse about time and motion seems circular, and he admits that time is a "hard" issue. Augustine reached the conclusion that time is essentially related to the mind, and not simply an ingredient of the inanimate physical world. Newton introduced "absolute, true, and mathematical" time that "flows equably without relation to anything external". Newton's mathematical (abstract) time is similar to our time (as an abstract entity), but his time "flows", while our time does not: our time is the bank (coordinate) in relation to which physical reality flows. Leibniz argued that Newton's absolute time (independent of everything else) does not exist. The only thing that exists and that we can know is the "order of succession" of events: there are no "moments" apart from events, argued Leibniz (Westphal and Levenson).

The two masters quarrelled (indirectly) about this issue which can be solved easily. Newton's mathematical (absolute) time can be considered the *abstract image* of Leibniz's order of succession in the case of an absolutely stable cyclic process. On the other hand, to speak about the order of events, we need a *definition (idea) of order*, such as the one given by Newton's description of mathematical time. In sum, the two discourses need each other, and together they lead to the same abstract entity which we call time. Newton simply made a step further from Leibniz's empirical "order of succession" of events to the *abstract representation* of that order, in mathematical form. The way to Newton's absolute (mathematical) time that "flows equably" runs through the experience of Leibniz's order of events.

23. It is generally considered that physics knows best what time is and what it looks like, because time plays a prominent role in its theories. However, physics consists of many theories; each of these theories describes a class of phenomena observed at a certain level of observation and from a certain point of view. For example, the theory of relativity describes physical reality at the level of massive bodies, while quantum theory describes the same reality at the subatomic level of description. Both these theories are successful at the operative level. However, "these two theories are known to be inconsistent with each other", so that "they cannot both be correct", says Stephen Hawking (p. 12). Such a claim is very general, but it points at the problem. Different theories often give different images of time, so that physics does not offer one description of time, but many. Hence, it is not possible to tell what "time" actually *means* in physics, if it means anything at all beyond the technical role it has in a specific theory. The word "time" is still intensely used in physics, but it has lost its old (simple, Newtonian) meaning, and the new meanings of this word are often vague or mutually incompatible.

24. At some levels of discourse and in some theories, time is assigned curious features. Time has been described as "fragmented" and "directionless"; it has been said that it does "not flow" and that it flows "in both directions", and so forth. Speaking about physical reality observed at the subatomic (quantum) level, Julius Fraser speaks about a "nonflowing, undirected, now-less time" (p. 106). Paul Davies says that observed at the level of quantum theory, "the universe has no well-defined time at all" (p. 181); Dieter Zeh speaks about the "timeless quantum world" (p. 197). Different levels of observation and discourse give different images of time, some of which look rather strange. This opens some essential questions, which we can only mention here. For example, when and how does time *emerge* in the physical world if it does not exist (or is not defined) at the subatomic level? Different theories speak about the same physical reality. Does it make sense to assume that there are several kinds of time in the physical world? We hold that it does not; in fact, we argue here that physical reality does not contain any time at all.

25. The basic assumption of physics is that "there is a real world out there that we can make sense of. And that world includes time", says Davies (p. 43). We do assume that the world is out there, although we cannot prove that this is really so. But we argue here that the world out there does *not* include time. Time is an element of the conceptual system by means of which the mind tries to make sense of what is out there; but time is not out there. *Change* is out there, but not time. Physics should try to base its discourse on the concepts of change (and of the intensity of change), which is immanent to the physical world, rather than on time, which does not exist in the physical world and which different theories describe in very different ways.

Let me repeat my mantra and sum up my image of time. Time is an element of the conceptual system by means of which people express their experiences, thoughts, and imagination. By means of this system (language) people describe their experience of existence in the changing physical reality in which they dwell. Time is not a physical entity (C1) and it does not exist in the physical world; it is an abstract entity created by the human mind (C2), and it belongs to the realm of abstract entities (C3). Time does not flow: it is an artificial bank (a coordinate) in relation to which physical reality flows (changes). Time is the means by which the mind *orders and quantifies* its experience of the world; temporal order can be projected on physical reality itself, but it is not part of that reality.

**4. The siren call of relativity**

26. The theory of relativity discarded Newton's absolute space and time, which are equal everywhere and for everybody: in this theory, the motion of each entity is observed only in relation to other entities. The theory of relativity was developed in two stages: the result of the first stage is called the special theory of relativity (STR), and the result of the second stage is called the general theory of relativity (GTR). STR speaks about the impact of speed on the flow of time and on space. GRT extends this discourse by including gravity into the story, which also influences the flow of time as well as space. Without entering into technical details (which are not needed here), we argue that the discourse of the theory of relativity is essentially a matter of *interpretation* of formulas and of their results, rather than of the formulas themselves. We argue that the relativistic interpretation of the formulas (of STR as well as of GTR) is inconsistent, and that it mixes basic ontological categories. The relativistic narrative speaks as if time were an ingredient of physical reality (C1); we argue that time is a creation of the human mind (C2), and an abstract entity that belongs to C3.

27. It has been said that theory of relativity is not only a matter of interpretation of Lorentz

transformation, but much more than that. This theory joined space and time into space-time and introduced a new, non-Euclidean, geometry of physical reality (space-time). This is correct, but it does not seem essential. Discourse about the relativity was introduced *before* the introduction of space-time and non-Euclidean geometry. STR was developed in 1905, and Hermann Minkowski developed the space-time formulation of STR in 1908. At that time, Einstein did not consider such formulation of his theory particularly interesting or useful (Lockwood, 52). It seems that the joining of space and time does not bring a substantial quality into the discourse and that by using space and time as separate entities it is possible to express everything that can be expressed by joining them together. However, several years later, Einstein "began to appreciate the strengths of Minkowski's approach", says Lockwood, so that he adopted the space-time model which then played an essential role in the shaping of GTR (p. 52). In a similar way, non-Euclidean geometry was introduced and adopted. However, in this entire story, only the Lorentz transformations (developed in 1904) seem of essential importance: they can be tested and they have been corroborated by numerous observations. The rest is basically a matter of formal (mathematical) shaping of the effects that these transformations revealed and predict.

28. Let us see some problems that the discourse about the relativity of space and time brings about. The claim routinely used by scientists and the media, that time flows slower with the increase of speed, is not as clear as it is taken to be. It is not clear (to me) whether time really flows slower, or it only seems so to a distant observer (of an entity) who is moving at a different speed. Discourse about this essential question is vague. A reviewer of one my paper tried to enlighten me with his explanation which consisted of expressions like the following one: "is *measured* as being shorter/slower than". I know that much: but what does this measuring actually mean? Is time flowing slower and space shrinking or not? Are they only measured as being shorter/slower, or are they *becoming* shorter and slower? I have not seen a clear and *consistent* answer to this question. We will assume here that the slowing down of time is considered *real*, because otherwise the discourse about the relativity of time would be only a peculiar interpretation of formulas and their results, and it would not tell much about reality. The next essential problem with relativity is how to tell *what* actually moves and at what speed in the world and in the theory in which there are no fixed space and universal time, and all movements are observed only as relative to each other. Incidentally, while writing this paragraph, the following question crossed my mind: Is there any difference between *being* and being measured *as being?* It depends on who measures, but I will not deal with this question here, because it is more complex than it may seem. But a proper answer to this question could resolve the difficulties of the relativistic narrative.

29. STR says that the time of the entity S (spaceship) that moves with the constant speed *v* in relation to the entity E (the earth), flows slower than the time of E, and that the magnitude of the slowing down of time depends on the relative speed *v*. Such discourse creates an obvious problem: S cannot move in relation to E without making E move in the same way in relation to S. This means that E has the same right and duty to claim that *its* time flows slower than the time of S. Therefore, STR produces claims that contradict each other, which means that it is inconsistent. The problem we described is called *twins paradox*, because it is illustrated by the example in which entities E and S are taken to be twins, let us call them Evelyn and Sandra, one of whom stays on the earth, while the other travels around in a spaceship. For many relativists, the discourse about the twins paradox is irritating, because they consider this problem solved and do not want to be bothered with this story any more. But the paradox is not solved and it cannot be solved in the context of the standard discourse of the theory of relativity.

Relativists claim that the paradox does not exit, because there is an "asymmetry" in the situations of the twins. To travel with the spaceship, Sandra must *accelerate*, which makes her situation different from the one of Evelyn who stays at home. We argue that without a fixed space, there is no asymmetry: Sandra in the spaceship cannot accelerate relative to Evelyn without making Evelyn accelerate in relation to Sandra with the same acceleration. The claim that the twin who "really accelerates" *feels* the acceleration does not help: STR does not register feelings. Furthermore, subatomic particles, which are used in experiments, do not seem to feel anything. Finally, if we assume (for the sake of argument) that acceleration does create an asymmetry, this does not solve the problem. Formulas of STR do not contain acceleration: they contain only the relative speed *v* between Sandra and Evelyn, which is the same for both twins. It is strange to slow down the time of Sandra on the basis of her *speed* relative to Evelyn (which is the same as the speed of Evelyn relative to Sandra), and then to justify the *choice* of Sandra by calling upon the fact that she accelerated. This does not look like a scientific explanation. STR does not give a coherent explanation *why* the time of Sandra (who accelerated) should flow slower instead of flowing faster. Acceleration is brought on the stage as a *deus ex machina* in the Greek and Roman drama, who decides what is and what is not. But acceleration is not a goddess.

30. "The twin paradox has generated an amazing amount of controversy over the years", says Michael Lockwood, and he offers the standard solution to this controversy. In brief, the *frame of reference* defined by the spaceship, "unlike that defined by the earth", is not an *inertial* frame, he says (p. 48), because the spaceship must accelerate to move away from the earth and decelerate to return to it. This is the standard argument based on acceleration. A frame of reference is called inertial if it is at rest, or it moves with a constant speed. Charles Stevens says that "an inertial frame is a coordinate system that is either fixed or moving linearly at a constant speed relative to the fixed stars" (p. 173). I am not happy with such definitions, because words such as "rest" and "fixed" mean nothing to me without a Newtonian fixed space. Lockwood takes it that the earth moves at a constant speed all the time, while the spaceship does not. However, it is *not possible* to know which framework is inertial and which accelerates, without a Newtonian fixed space that is inertial by definition.

Acceleration means change of velocity, "which includes deceleration and change of direction", explains Lockwood (p. 49). We expect acceleration at the take-off and at the landing of the spaceship carrying Sandra, but these accelerations can be eliminated from the story. We can assume that Sandra simply *passes* with her spaceship near the earth at the beginning and at the end of a period of ten years (earth's time) and that the twins compare the states of their clocks as they pass by each other. In this way, the take-off acceleration and the landing deceleration are removed from the story. How can we know *whose* time flows slower in such case, and avoid the paradox? To answer this question, Lockwood calls upon "the acceleration that the spacecraft undergoes at the turn around stage" (p. 49). To pass twice near Evelyn on the earth (at the beginning and at the end of the experiment), Sandra in the spaceship must turn around somewhere in space, and "this acceleration at the turnaround" is the one that "introduces a crucial asymmetry into the situation, which is the source of the difference of ages at the end" (p. 49). It seems incredible to me that something like this can be offered and accepted as an argument in scientific discourse.

31. Let us see the above argument one more time. The fact that Sandra had to turn her spaceship around, so as to pass by her twin sister Evelyn on the earth (to compare their clocks), has been given as the *reason* for Sandra's time to flow slower as long as she travels (possibly forever), including the period *before* she turned her spaceship around! "That sounds like magic to me, not science", said Hawking on some other occasion (Hawking and Penrose, 124); the same should be said for the incredible effects of the "turnaround acceleration". The claim that the turning around of the spaceship has the power to affect not only the future, but also the past, is magic of the highest degree. This magical turnaround acceleration is the *only* thing that "protects" the discourse about the relativity of time from precipitating into the abyss of inconsistency and meaningless discourse. But this turnaround acceleration cannot be taken seriously, so that we consider the standard discourse about the relativity of time inconsistent and meaningless.

Let us add to this that Lockwood's argument is inconsistent, too. He takes that the frame of reference defined by the earth is inertial, but this is obviously not so. A frame of reference is inertial if it is at rest or it moves with a constant speed: therefore, it must not accelerate. Lockwood defined acceleration (above) as a change of speed or direction. The earth rotates around the sun, which means that it *constantly* accelerates ("turns around"), not only once, like Sandra's spaceship when it turns around to return to the earth. This shows that the earth and the spaceship are in a basically equivalent position, which is understandable since there is no fixed space.

32. After many years of arguing that acceleration does not solve the twins paradox, I found an author, Tim Maudlin, who accepts the theory of relativity, but considers the argument based on acceleration invalid. Maudlin shows a couple of invalid arguments which claim that acceleration solves the twins paradox. One of these examples comes from "the great Richard Feynman", who presented this argument in his trilogy *The Feynman Lectures on Physics*. Maudlin puts forward a long quotation from Feynman's discourse about how acceleration solves (eliminates) the paradox. But at the end, Maudlin informs us that "everything in this 'explanation' is wrong" (p. 81). This is what I have been saying for many years. Maudlin also shows that it is easy to create a situation in which the twin on the earth accelerates *more* than the one that travels by spaceship, "but still ends up older" than the twin who travels by spaceship (p. 81). The case with acceleration shows that also great physicists can fail to see trivial mistakes, and be dogmatic, too.

33. "Special Relativity is a very simple theory that is commonly presented in a complex and confusing way", says Maudlin (p. 67). Einstein presented the theory as something that follows from two basic principles: (1) the equivalence of all inertial frames, and (2) the constancy of the speed of light. "From these two principles ... we derive the *Lorentz transformations*, which are a set of equations relating one set of coordinates to another" (p. 67). Maudlin holds that this way of presenting the theory is not appropriate. First, "the motion of an inertial system, or an inertial set of coordinates, or an inertial frame of reference, is derivative rather than fundamental" (p. 67). Such concepts "can only be defined by reference to some objective geometrical structure of space-time itself, in order to make sense of the qualifier 'inertial'" (p. 67). This is basically correct, but the qualification of geometrical structure as "objective" seems problematic. I am not sure that there exists such a geometrical structure, nor whether such a structure can exist. Maudlin holds that "we ought to begin with the intrinsic geometry, not with coordinate systems or reference frames" (p. 67). I accept such an approach, but again, the discourse about "intrinsic geometry" seems too strong. It would be nice to know such a geometry, but it is not sure that such a geometry exists nor that it could be known (if it existed). Anyway, Maudlin holds that "Special Relativity is, fundamentally, a postulate about the structure of space-time" (p. 83).

34. The second problematic concept in the presentation of relativity is speed. "Basing a presentation of Relativity on talk of speeds unavoidably suggests that we are dealing again with Newtonian absolute space and time and Newtonian absolute motion", says Maudlin (p. 67). I completely agree. Furthermore, popular presentations of the theory of relativity say that with the increase of speed of an object its time runs slower and that its space shrinks. Such claims imply that objects *have* specific speeds. "But in Relativity ... there simply are no such speeds", says Maudlin, so that to understand the theory of relativity properly, "we have to expunge all ideas of things having speed, including light" (p. 68). I completely agree; I speak about speed because this is how the theory of relativity is normally presented, but I agree and I said it above that such discourse is obviously problematic: it is either Newtonian or ungrounded. "Instead of trying to somehow derive Special Relativity from some phenomena or some general principles, we will simply state, in as clear a way as possible, what sort of space-time structure Special Relativity postulates to exist", says Maudlin (p. 69). We can then consider (describe) various physical phenomena in terms of that geometry, and see whether the outcomes of such considerations seem correct or not.

35. Points in the space-time are called *events*, and clocks are "like odometers on cars, measuring the length of their trajectory through space-time", says Maudlin. "Clocks that wander off on different trajectories can record quite different elapsed times between the same pair of events, just as cars that take different routes between the same locations can show different elapsed miles on their odometers" (p. 76). In the context of his geometrical presentation of the special theory of relativity, Maudlin has a simple solution to the "iconic" twins paradox. In order to determine how much time has "elapsed" for each of the twins, we need only to compute, for each of them, the trajectory she made in the Minkowski space-time from the event (point in space-time) of their separation to the event (point in space-time) of their reunion. According to Maudlin's geometrical presentation of the special relativity, the trajectory ("world-line") in the Minkowski space-time of the twin that remained at home is *longer* than the trajectory of the twin that travelled by spaceship and then returned home. This means that the clock of the twin that travelled shows less time, and that this twin has now become younger than the other one. In Minkowski space-time, "clocks measure the Interval along their world-line", and the world-line of the twin at home is longer than the one of the twin that travelled. Hence, the one that travelled is now younger (Maudlin. 79).

I cannot spend much space-time on showing that Maudlin's simple solution of the twins paradox might not be as simple as he claims it is. He stresses that the paradox has been solved "without having to attribute any 'motion' or 'speed' or 'rest' to anyone: it is a simple matter of space-time geometry" (p. 83). It may look so, but it seems to me that the original problem remains present in Maudlin's discourse. Namely, how do we *know* which of the twins travelled? Minkowski space-time is a formal system which allows us to compute various trajectories (world-lines) in that system. But this is not enough: we need a criterion on the basis of which we can differentiate the relative mutual movement of the twins. "Minkowski space-time does not support any objective measure of the speed of anything", says Maudlin (p. 95). "Once we abandon Newtonian absolute time and the persistence of points of Newtonian absolute space, there are no objective speeds, either of light or of anything else" (Maudlin, 121). I completely agree. But how can we know which of the twins "really travels" or along which world-line a twin moves? Maudlin does not say such things, and it seems to me that there is a problem here, but I do not have spare space-time to analyse Maudlin's discourse further.

36. Speaking about the STR, Jean-Pierre Luminet describes the notorious case of particles called muons. It has been shown that by the increase of speed of these particles their decay half-time, "as measured in the rest frame of the laboratory", increases in accordance with the formulas of STR (p. 529). Such results corroborate formulas of STR (Lorentz transformations), but calling upon "the rest frame of the laboratory" makes the discourse actually Newtonian. Without the assumption that *muons* are the ones that move and the laboratory is at rest, the discourse about slowing down of time for muons would be in trouble, as the twins paradox shows. Finally, muons are processes (they decay), as everything else is: with the increase of speed with which muons move, these *processes* slow down, and hence their lifetime increases. That is all and this is enough. To interpret this *fact* by the claim that *time flows slower* for them, sounds exciting but it brings serious difficulties (contradiction), and it does not seem (to me) particularly useful.

The measurements that confirm the validity of formulas of STR do not confirm directly (by themselves) the interpretation of these formulas given by STR. Formulas can be tested, but the interpretation of formulas cannot be tested and confirmed in the simple way in which formulas usually can. An interpretation of a correct formula can be wrong, logically and factually. We can accept empirical facts that with the increase of speed, muons live longer, clocks tick slower, and so forth, but this does not mean that for these entities *time* flows slower. The fact that processes slow down in some conditions does not mean that time slows down; on the other hand, the assumption that time slows down creates problems which we have not been able to solve, and we may never be able to solve in a coherent way. Tricks with the "acceleration at turnaround" do not help: they only show how hopeless the standard discourse about the relativity of time is.

37. Motion (speed) may slow down *processes* (clocks included), but this does not mean that it slows down *time*. Measurements do not show the speed of time: we measure (compare) the speed of processes, not the speed of time. Every measurement is a comparison of two entities of the same kind. Nothing can show the slowing down of time, because nothing measures time: time does not exist in the physical world, so that it cannot be measured. However, if all processes in an entity or space slow down, can we then not say (assume) that time itself slowed down for that entity or in that space? No, because time is not a physical entity: it does not flow, and hence, it cannot slow down. No, because time is the measure by means of which we measure the amount and intensity of change: if we make the meter relative ("malleable" and "flexible"), then the values we measure lose a clear meaning. Or perhaps *we can*, if we really wish to do it. However, the inconsistency of the discourse about the relativity of time shows that we have not discovered yet how to do this in a consistent (meaningful) way.

Steven Savitt says that the empirical evidence which corroborates the validity of STR is overwhelming, and adds that when "the evidence supports a theory that forces us to an odd conclusion, common sense must bow to the evidence" (p. 565). This is an unfortunate claim, because Savitt does not make a distinction between the Lorentz transformations and their relativistic interpretation. We can bow to the former, but we have no reason to do this to the latter.

38. I once wrote that the theory of relativity is the shifting sand in which many people have tried to swim during more than a century, but they have not advanced much. If we do not have a better interpretation of the Lorentz transformations and their effects (results) than the standard relativistic narrative, then we do not have a satisfactory interpretation of these basic formulas, and we should face this fact. On the other hand, I admit that the relativistic narrative has its unique charm. The assumption that there is a local time *t* at every point of the universe *p* for every observer *o* is amazing, indeed. Such assumption fragments reality into countless *solipsistic realities* or into countless *images* of the one reality: for each observer, his own unique reality (image) with his own local times at local places. This looks amazing, indeed. Is it not the reality of each person actually exactly such, mentally and physically? I must stop dealing with relativity, before it takes me away by its siren song.

The assumption (postulation) of anything absolute seems too strong, but we need a firm framework of discourse: a coordinate system that facilitates a precise discourse about the reality that we perceive. However, the adoption of such a coordinate system is problematic, because there is no specific point in the universe, which could be considered a natural centre (origin) of such a coordinate system. We can place its centre wherever, because there are no points in the space of existence by itself, before a coordinate system is introduced. Hence, it may be natural to consider *all points* equivalent and equally worthy to be centres of their own coordinate systems. In this way, the relativity of space and time can be introduced and justified. Is this my surrender in front of the siren call of relativity? Not exactly. The relativity of space and time has its charm and appeal, but the standard discourse about relativity is far from being satisfactory, and it must be changed even if the basic assumption is retained. Furthermore, I am not sure that it is useful to introduce infinitely many coordinate systems (frames of references) and infinitely many times, as the STR has done. I am also not convinced that it is useful to join space and time. Leaving technical issues aside, my basic position is that time and space are a *matter of language, not part of physical reality*, as relativists argue and assume.

**5. Crawling through eternity**

39. GTR adopted a non-Euclidean geometry and introduced a four-dimensional space of discourse about physical reality, in which the traditional space and time are joined into space-time (or spacetime). A point in space at a point in time (that is, a point in space-time) is called a *world point* or *event;* the totality of all world points forms the (four-dimensional) *world*. The existence of an entity (that we perceive as enduring in time) is presented with a curve in the four-dimensional space, called *world line*. Points of such a curve can be labelled by their three spatial parameters and by the successive values of the parameter *t* associated with the clock carried by the entity (which shows the *proper time* of the entity). In space-time, events do not happen: they always *are;* the illusion of happening (change) is created by the movement of the observer in that realm of existence. "The objective world simply is, it does not happen", runs the notorious statement attributed to Hermann Weyl. Only to the "gaze" of a consciousness "crawling upward" along the world line of the body (from which it emerges), "does a section of the world come to life as a fleeting image in space which continuously changes in time" (in Whitrow 1980, 348). Lockwood says (p. 54) that this description was given by Arthur Eddington and that it was mistakenly attributed to Weyl, but this is not important here. What matters is whether such discourse is clear and consistent: we argue that it is neither of the two.

40. Space-time and GTR have been accompanied (interpreted) by the *block universe* view of existence. This view (model) assumes (argues) that reality is a four-dimensional space and that it contains all entities and events. Nothing ever becomes and nothing ever vanishes in the block universe: everything there is has always been and it will always be, so that this view is called *eternalism*. What we consider the past or the future moments, entities, and events, exist in the same way as those we consider present: they are only not temporally here. We can understand that entities which are not spatially here, do exist: in the same way, we must understand that entities which are not temporally here, do exist. This analogy is the strongest argument in support of eternalism, but this is an argument by analogy (at best), and such arguments are generally weak. There are other attempts to show that the block universe view is plausible and coherent, but they are even weaker than this one.

It has been said that the block universe view gives a *static* image or understanding of reality (existence), but Huw Price points out that such a discourse is not appropriate. The block universe is not "an entity *in* time", so that it cannot be described either as dynamic or as static, because such descriptions can be applied only to the entities which are "*in* time". Time is one of the four dimensions of the block universe, because this universe contains all "moments" (or "versions") of every entity that *is* (has ever "been" or will ever "be"). Time does not flow in this universe and there is no "privileged now" in such a universe (p. 13). Every entity and event is always there (in space-time), regardless of when it enters into the experience of an observer.

41. Descriptions of the block universe view say that entities do not appear and vanish, and events do not happen: they *always are* in the four-dimensional realm of existence, called spacetime. Happening is an illusion created by the consciousness that "crawls" along the world line of the body from which it emerges. However, it seems that the discourse about "crawling" contradicts the claim that everything "is" and nothing "happens". Is not this crawling a happening? In what sense and way does an entity crawl if everything always *is* and nothing ever *happens?* Why does a consciousness crawl only for a limited amount of space-time, if everything always is and never vanishes? The block universe view has a mystical charm and the appeal of a fairy tale, but it lacks the clarity, precision, and consistency of scientific discourse. This view is also in a complete discrepancy with our perception of reality which seems in constant change at all levels of observation. Hence, contrary to Weyl and Eddington, I say: *the objective world is not: it is constantly becoming and vanishing*. Existence is a process: nothing is; everything is becoming and vanishing, including me, the ephemeral observer of this ceaseless game.

42. In the block universe, all entities and events that we consider past, present or future, are equally real, says Davies: all people, past, present, and future are "there" (p. 260). This world looks rather crowded to me. "The entire universe exists as a single block with no parts, so no part of it either comes into existence or goes out of existence", explains Raju (p. 257). This is a mystical discourse and I do not understand it. Such explanations do not tell me anything, because they do not speak in an understandable way. Has this sentence always been written? If yes, what am I doing here and now, while I am writing it? Am I only "crawling" from the world-point-1 (in the space-time) in which this sentence is not, to the world-point-2 in which it is? But why must I type on the keyboard of my computer to move from one world point to the other? The block universe narrative seems too vague, incomplete, and incoherent to be analysed in any precise way.

"The most natural way of thinking of time, in the context of contemporary physics and cosmology, is to regard it as one dimension of a multidimensional space-time manifold", says Lockwood. Furthermore, "our best theories" support the block universe interpretation of GTR as well as of existence (p. 249). However, Lockwood admits that the block universe view "remains a controversial view, which may conceivably turn out to be mistaken" (p. 249). Let me remind you that this view is an interpretation, which cannot be empirically tested in a direct way. This interpretation is very vague and incomplete, so that it is not clear *in what way* could it be tested and shown either correct or mistaken at the level of empirical facts. Relativists often emphasise that Einstein personally supported the block universe interpretation of GTR (Zeh, 195), but a personal choice is not a scientific argument. In sum, the block universe view belongs to poetry rather than to science.

43. The introduction of space-time is a formal (mathematical) issue, but it changed the discourse and the way of understanding space and time. Newton introduced absolute space and time as the conceptual means that facilitate a precise discourse about physical reality. His space and time serve for "keeping track of motion mathematically", but they do not "*do* anything", says Davies. On the other hand, Einstein "restored time ... at the heart of nature, as an integral part of the physical world" (p. 17). In the theory of relativity, "space and time ... play a full and active role in the great drama of nature", says Davies (p. 16). This sounds exciting, but it is wrong to assume that time is part of the physical world and that it *does* anything. Relativists assume (and argue) that "space and time ... are *physical* things, mutable and malleable, and, no less than matter, subject to physical law" (Davies, 16). These quotations put forward the essence of what I consider mistaken in the discourse about GTR and its block universe interpretation. Space and time do not play any role in the drama of nature: they are not physical things, mutable and malleable, and they are not subject to physical laws. Space and time are *abstract* entities (categories), created by the human mind (C2), and they exist only in the realm of abstract entities (C3). Space and time facilitate precise *discourse* about physical reality; they are *not part* of physical reality.

44. Lockwood says basically the same as Davies. He describes the theory of relativity as a momentous revolution in thought, which made "space and time, as aspects of space-time, active participants in physical interactions" (p. 79). The old conception of space as "a merely passive arena" in which forces play their games, and of time as "merely the road along which the march of history proceeds", has given way to "an actively involved space-time", which "feels" the matter and energy, and "kicks back" accordingly (p. 79). This sounds exciting, but it says (explans) nothing, because this is a play with nice metaphors, not a concrete discourse.

In Newtonian physics, space and time "serve as the stage on which the history of the physical universe plays out", says Maudlin (p. xii). Relativists often say that Newtonian space and time form the "arena" or the "stage" in/on which physical events unfold. But such claims are not appropriate. Newton's space and time are the abstract means by which we describe physical reality as a process that unfolds *in nothingness*, because there is nothing beyond it. Time and space are not the stage: physical reality (the universe) does not need any stage to be what it is. We, the people, need *concepts* to be able to speak about physical reality, and to describe what we perceive and experience. Time and space are parts of our language. By mixing basic ontological categories, such as physical reality (C1) and language (C3), it is possible to produce many nice metaphors and stories which do not explain anything and do not say anything substantial.

45. In the block universe (space-time) everything always *is;* hence, it is assumed (and argued) that this space of existence allows *time travelling*. This means that one can travel into those regions of this realm, which we call "past" and into those that we call "future". I consider the discourse about time travelling vague and inconsistent, so that I would rather not spend much space-time on this issue. The discourse about time travel leads to various problems. The best known among them is the case of a grandson who travels to the past with the intention to kill his grandfather at the time when he was young. If the grandson does this in the time before the grandfather had children, then the grandson should not have been born at all. And if he was not born, how did he travel back in time and kill his young grandfather? Let us mention that there is no need to travel that far to create the paradox: it is enough to visit and eliminate your father before he had children. Sigmund Freud would have preferred the latter option, but it seems that relativists prefer the grandfather scenario.

To avoid paradoxical situations (inconsistencies) to which time travel leads, it is assumed that the physical world contains some "insuperable physical obstacles" which do not allow those events that would create a paradoxical (inconsistent) situation, such as the one with the premature killing of one's own grandfather, to take place. But there is no need to deal with "physical obstacles" here: the fact that an assumption leads to inconsistency is a *logical* obstacle strong enough to prevent us from building a rational discourse on such assumptions. Paradoxical implications of the block universe interpretation of GTR, which facilitates time travelling, mean that this interpretation should better be abandoned. But instead of doing this, relativists expect the physical reality to protect itself and us from the unsavoury consequences of our wrong assumptions, theories, and interpretations. Difficulties with time travelling come from the wrong assumption that the past and the future physically exist. Physical reality is a river that flows from non-existence into non-existence. There is no basis for the assumption that the twenty year old "me" exists and lives somewhere, and that he could be visited and even killed. Only the present "me" exists: the one of twenty years ago has gone and is no more, as well as the one of twenty seconds ago. I am a process, as everything else is. It is not possible to travel to the past or to the future because such places do not exist.

46. Davies considers Einstein's discourse about space and time "a monumental first step", but he admits that this revolution has remained "frustratingly unfinished" (p. 17). I consider this "first step" a mistake because it mixes basic ontological categories (the physical and the abstract) so that this revolution will remain unfinished for the foreseeable future and much longer, probably forever. I got the impression that the discourse about relativity began as a joke that nobody took seriously, but everybody considered charming. Anyway, it seems that this revolution, which started *after* Lorentz developed his transformations, has not produced anything particularly useful, and that the Lorentz transformations are the only thing that really matters and functions well in this entire story. On the other hand, the discourse about the relativity of space and time seems dubious, to say the least. Einstein "triggered a revolution in our understanding of the subject, but the consequences have yet to be fully worked out", says Davies (p. 9). The difficulties (paradoxes) we put forward above indicate that the present state of the relativistic narrative is not encouraging. "We are still a long way from solving the riddle of time", says Davies (p. 280), but he holds that we are on the right path towards the solution: "we must embrace Einstein's ideas, but move on", he says (p. 10). However, if we embrace an ontologically and structurally wrong discourse we will not reach far. I hold that the assumption that space and time are "physical things" and "malleable" will not lead us further than to paradoxes and contradiction.

47. When I write about the theory of relativity, I wonder how much of this story I understand in a proper way. I surely do not understand everything, and probably nobody else does. Because "Einstein left things in a curiously unfinished state", says Davies (p. 16), an advocate of the theory of relativity. "The revolution begun by Einstein remains frustratingly unfinished", he admits (p. 17). Therefore, it seems that nobody should claim that he or she understands this frustrating and unfinished story completely. Einstein lived fifty years after he published his STR: it seems strange that he had not done more on the "finishing" of the revolution he triggered, assuming this *can* be done, which I am not convinced it can. I have an impression that the discourse about the relativity of time and space is a charming game which does not say anything relevant and useful, beyond the Lorentz transformations which were developed before the theory of relativity, and which can work well without the frustrating relativist burden that has been thrown on them. Davies warns readers of his book about relativity that after they read his book they could be confused even more than they were before they read it. But "that's all right", he says: "I was more confused myself after writing it" (p. 10). I can understand him.

## 6. Concluding remarks

48. Time is a conceptual rather than an empirical issue: it is a matter of ontology and logic rather than a matter of experimental physics and testing. We must *define* time and its features rather than discover them in the physical world, and we must do this in a consistent way. Physical reality is a process: it *is* as a ceaseless becoming and vanishing. Change is immanent to physical reality: the experience of change is the source of the idea and concept of time. Time is not part of physical reality and it does not flow, either absolutely or relatively, either forward or backward. Time is the artificial bank in relation to which physical reality flows. Time belongs to language: it is one of the basic concepts in terms of which we describe physical reality and the way it changes.

49. Physical reality is a process that takes place in nothingness: it simply *is as a happening*, as a ceaseless becoming and vanishing. Change is immanent to physical reality: it is ontologically and epistemologically prior to time; the experience of change is the source of the idea and concept of time. The fact that speed and gravity "slow down clocks" does not mean that they slow down time. Physical entities, clocks included, are processes: various forces can slow down processes. However, there is no need to make the leap from this fact to its *interpretation* which says that this means that time flows slower for these entities. Processes can slow down, but time does not and cannot, because it does not flow. It is necessary to differentiate physical reality and abstract entities (language) by means of which we describe this reality. Time is a metaphysical (linguistic) category, rather than physical: we must define its features, rather than discover them.

50. Discourse about the relativity of time is essentially a matter of interpretation of formulas and the results they produce, not a matter of testing the formulas. Empirical observations that confirm the correctness of a formula do not, by themselves, confirm the correctness of any of its possible interpretations. A formula that describes correctly a quantifiable behaviour of a class of phenomena can be interpreted in a formally inconsistent and factually wrong way. In principle, we can interpret formulas and their results as we wish, but if an interpretation implies a contradiction (paradox), then there is something wrong with it. Logical consistency is the minimal request that every theory and interpretation (explanation) must satisfy to be understandable and to be at least possibly correct. When we do not have a satisfactory explanation of some formulas that "work well", we are inclined to embrace any explanation that sounds interesting or appealing. But in such a case we should not claim that we have the explanation (consistent and complete), because we do not.

51. The block universe view, also called *eternalism*, assumes that "there is no ontological distinction between past, present and future, and that all times are equally real" (Dainton, 57). Such a view is exciting and appealing, but it looks to me like an initial idea for the production of fairy tales, rather than a plausible scientific discourse. On the other hand, the view called *presentism* claims that only the present exists: the past is no more, and the future is not yet. Presentists "deny any reality to the past or future", says Dainton (p. 57). "Presentists hold that only the present exists ... whereas eternalists grant equal reality to all times", says Le Poidevin (2015, p. 8). Eternalists are truly magnanimous towards all beings, alive and non-alive. Anyway, I consider presentism trivially correct, but I am aware that it looks too simple to be accepted by people who love charming fairy tales. It has been pointed out that we actually perceive the past, not the present, because signals (light and others) need a certain amount of time to arrive from the observed object to the senses of the observer. This time is usually very small, but from distant stars, light can travel for many years. However, this is not relevant for the discourse about presentism and eternalism. Presentism claims that only the present exists (even if seen with a delay) and that existence is a process of ceaseless becoming and vanishing. Eternalism tells the opposite story: everything always is, and nothing ever becomes, passes or vanishes. Between presentism and eternalism, there are other models of existence, which adopt some elements of these models, leave out others, and introduce some new elements, but we do not need to deal with these models here.

52. Existence is a happening (becoming and vanishing) that the mind shapes into events. Past and future do not exist in the physical sense. "Past" is the name for those things that existed and vanished, and do not exist any more, except as our memories and records. "Future" is the name for things that do not exist, except as our plans and expectations that may materialize or not, depending on various factors. It is not possible to travel to the past or to the future, because such places do not exist. It is not possible to return "back in time" or to "jump forward in time", because such places do not exist. A theory that supports speculations about such jumps and travelling is suspicious, to say the least. "To exist" and "to exist now" means the same.

We cannot perceive future events, because there are no future events. There are only present events, and our expectations of the future present events, or simply the expectation that *new* events will take place. We cannot perceive past events, because there are no past events: there are only the present memories and records of the past events. Only the present is: the past is the name for something that was, but is no more. The future is the name for something that we expect to be, but is not yet. It has been said that "the past is fixed" and that it is not possible to "affect the past"; but the past is not simply fixed: the past does not exist any more, and that is why it is not possible to affect it. On the other hand, it *is* possible to affect the future, because we belong among the forces that make impact on the process of existence *now*, and with this we shape our future presents.

53. Presentism has conceptual problems. Strictly speaking, the present or "now" must not have any duration (temporal extension). Because if the "now" had any duration, some of its parts would come earlier, other later, so that such a "now" would be a time interval, not the present. Hence, "now" must be a durationless interface between past and future. The present ("now") is a gate through which reality enters into existence, and through which it immediately vanishes. How can we perceive something that has no duration, and how can we do this in a durationless present? How can the mind, that emerges from physical processes, perceive (register) anything that is durationless (without extension)? Furthermore, how does human mind create the sense of duration and movement, if all that it perceives are durationless "nows" (one after another)? I do not know impeccable answers to these question; our language and cognitive models have not been able to describe many things in impeccable ways. In this case, it has been assumed that the senses of duration and motion are something that the mind constructs from perceptions, memory, and expectations (Le Poidevin 2015, 3). This allows conscious beings to envisage and understand the world (process) in which they live and of which they are part.

54. Concepts such as the laws of physics, discovery, and similar are often used in inappropriate ways. People routinely speak about Einstein's "discovery" that there is no universal time and that time is relative, as if time were an island in the ocean, which was discovered. The relativity of time was not discovered, but assumed: this was done as an interpretation of the Lorentz transformations. Physical reality contains no time the property of which could be discovered.

Fraser speaks about "Einstein's discovery that $c$ constituted a natural upper boundary to the speed at which energy (and hence messages) may travel" (p. 241). But something like this *cannot* be discovered. The value $c$ can be the upper boundary of a variable in a given theory, but we can never know whether there is a higher speed in the universe. We cannot know what the largest value in reality is, because it is always possible that a larger value will be discovered tomorrow. A new theory can be developed, with new concepts (ontology), laws, and "upper boundaries". It is not possible to discover upper boundaries in physical reality: only a maximal value of a variable in a given theory can be known. But every theory is only a hypothesis which can be displaced and replaced by a new theory.

55. Physical laws do not "govern" processes in the physical world. These laws are our descriptions of these processes which take place by themselves and are not governed by anybody and anything. It is often said that the world is "governed" by *time-symmetric* laws, but the world is not "governed" by anything. The world simply *is*, and it is such as it is, and it evolves (functions) as it evolves. It is not governed by anybody and anything. Laws of nature are created by people and they are part of our language, by means of which we describe our perception and understanding of reality in the best way we can. Some assign a big importance to the fact that laws of physics "allow" processes to evolve in *both* "temporal directions" equally, but processes evolve in *one* direction only. They really do, but so what? Physical reality is a process that evolves in the way it does. There is nothing unusual in all this: it is simply so. On the other hand, the laws of physics are a creation of the human mind and they are such as we were able to make them. The fact that these laws allow to happen also those things that do not happen shows that they are imperfect as we ourselves are. If our laws are "time-symmetric" and physical reality is not, we should try to produce better laws.

56. Some speak about time turning back and running in the opposite direction. But such

discourse is vague and semantically dubious. First of all, there is no time in the physical world, so that it does not run anywhere, neither forward nor backward. If some processes were to begin to evolve in the way that is opposite to the one they normally evolve, this would be a change in the way these *processes* evolve, not a change of the direction of time. If in ten year's time I were to look twenty years younger than now, this would mean that the process called my body has evolved in an unusual way. It could be said that the process called my body had "turned back", but it would make no sense to say that time (for me) changed its direction. Time does not flow, so that it cannot flow backward: a process can, in principle, take place in an unexpected way, for reasons that may not be known to us. If physics one day tells us that it managed to turn time back, but my mirror (pain in my knee, etc.) tells me that I am still getting older, then this turning back of time will mean nothing to me.

57. Wittgenstein is considered the most sharp-minded author of the twentieth century in the space of language and knowledge. I have not managed to find much substance in his writings, but some cryptic claims of this mystical soul support my basic view that time is not an ingredient of the physical reality. In *Tractatus* (6.3611) Wittgenstein says: "We cannot compare any process with the 'passage of time' - there is no such thing - but only with another process (say, with the movement of the chronometer)". Therefore, "there is no such thing" says Wittgenstein about the passage of time, and he is right. We cannot compare anything with the "passage of time", because there is no such thing as the passage of time. We compare the duration of processes: we measure time by means of a chosen cyclic process: a unit of time is a certain number of cycles. The clock is an observable physical device that assigns numbers to cyclic processes.

58. Lot of things have been said about the beginning of time. We cannot deal with this issue here; let us only mention that time should be considered infinite for semantic and logical reasons. The concept of beginning can have a meaning only *in* time, so that we cannot speak about the beginning of time itself in a coherent way. Time did not begin with the Big Bang, because the Big Bang was not the absolute beginning. Every "bang" is a process, and every process is caused by another process. It is not possible to imagine the absolute beginning of everything or to say anything rational about such a beginning. An absolute beginning would mean that *something* (everything) came out of *nothing:* such an event transcends the human understanding, and probably the divine understanding, too. Such event would transgress the basic laws of physics, as well as of the human thinking. No cognitively relevant explanation about the origin of existence can be given. A discourse about the end of time is equally problematic for semantic and logical reasons.

59. It has been said that we should be prepared for "a new structure for the foundations of physics that does away with time", because time is in trouble (in Davies, 178). The sophisticated expression "a new structure for the foundations of physics" means a new metaphysics or ontology. Therefore, a new (or changed) set of basic concepts (classes, entities) could be introduced, together with their descriptions, by means of which physics would describe reality and events, and shape our perception, cognition, and imagination. The discourse about time in physics has seemed to be in trouble for a long time, and it may be that the time for changes has come. Physics will abandon the discourse about the flow of time one day and replace it with the discourse about the *intensity of change*. But people will keep the plain old time for their reflections on existence and ephemerality, because although time does not pass, everything else does.

60. These were my reflections about time. They are surely not impeccable, but they may not be completely worthless, either. Speaking about the difficulties related to the applying of quantum mechanics to gravitation, Davies says that physicists are "divided about the need to pin down a sort of 'master' time, a natural measure of change in a physically uncertain world, or to define time completely out of existence" (p. 281). I love the expression "a master time, a natural measure of change": this is what I have been preaching about for many years. I would not call this "measure of change" natural, but artificial because all measures are human creations. Regarding the above dilemma, I would happily adopt *both* options: therefore, (1) time as a measure of change, which (2) does not exist in physical reality, because it is a creation of the human mind.

My favourite ideological ally, Lucretius, a Roman thinker and poet who lived in the first century, claimed that "time cannot itself exist, but from the flight of things we get a sense of time" (in Davies, 23). Therefore, he considered time a creation of the human mind, not a physical entity. "Could it be that, after millennia of deliberation about time, we shall finally discover that it doesn't really exist as a basic ingredient of reality", asks Davies at the end of his book (p. 281) in which he argues the opposite. It is high time for physics, indeed, to make this "monumental first step"; the wise Lucretius did it two thousand years ago. It seems that the cryptic Wittgenstein discovered the secret, too, and I have always known it.

+